# MEMS for Photonic Integrated Circuits

Carlos Errando-Herranz, Alain Yuji Takabayashi, Pierre Edinger, Hamed Sattari,
Kristinn B. Gylfason, *Member, IEEE* and Niels Quack, *Senior Member, IEEE*

*Abstract*—The field of microelectromechanical Systems (MEMS) for photonic integrated circuits (PICs) is reviewed. This field leverages mechanics at the nanometer to micrometer scale to improve existing components and introduce novel functionalities in PICs. This review covers the MEMS actuation principles and the mechanical tuning mechanisms for integrated photonics. The state of the art of MEMS tunable components in PICs is quantitatively reviewed and critically assessed with respect to suitability for large-scale integration in existing PIC technology platforms. MEMS provide a powerful approach to overcome current limitations in PIC technologies and to enable a new design dimension with a wide range of applications.

## I. INTRODUCTION

Photonic Integrated Circuits (PICs) bring the benefits of miniaturization to optics, promising leaps in performance and scalability, as well as a dramatic reduction in cost and power consumption for a wide range of optical systems. The numerous applications of PICs, including high-speed telecommunications [1], high-performance computing [2], label-free biosensing [3], and quantum technologies [4], have resulted in a rapidly increasing market size and research interest.

Common PIC technology platforms use silicon, III-V semiconductors, or silicon nitride as waveguiding material. Silicon is one of the most promising PIC material platforms for large-scale integration, due to its high-quality fabrication processes, heritage from the microelectronics industry, and its high refractive index. Consequently, silicon PIC foundry processes have been introduced in the past decade [5]–[7], providing standard photonic library components [8] and easy access [9] including prototyping runs using multi-project wafers (MPW) [10], [11]. This approach drastically reduced the cost of PIC development, making state-of-the-art technology accessible to industry and academia alike, which has resulted in several large-scale PIC demonstrations in recent years [12]–[18].

While these breakthroughs have already accelerated the development and market introduction of PIC-based products, technology is constantly pushed forward to improve performance and add value. In order to bring PICs to an even larger scale, efficient tuning mechanisms are required, to compensate for manufacturing variations and environmental perturbations, or to enable reconfiguration. In this quest for enhancing current PIC technologies, the standard platforms are constantly augmented, e.g. by introducing new materials or process modules, while typically ensuring full compatibility with the existing platform.

A promising route for enhancing current PICs is the exploitation of mechanics at the nano- and microscale. MEMS in PICs provide: (1) an efficient tuning mechanism to adjust the operation point of photonic components for large-scale PICs, (2) enhancement of current capabilities (e.g. by introducing bistability and making zero-power consumption states possible), and (3) entirely new capabilities, such as mechanical motion for dynamic coupling optimization to fiber interface. MEMS is a mature technology based on semiconductor manufacturing techniques and has been successfully integrated in numerous high-volume products such as accelerometers, gyroscopes, and micromirror arrays. The integration of MEMS in PICs is thus a natural extension, and several implementations of MEMS in photonics have successfully been demonstrated.

Previous reviews on photonic MEMS have focused on non-integrated, free-space optical MEMS [19], [20], or on individual devices [21]–[23], MEMS based optical cross connects [24], and optomechanics [25].

Here, we review the field of MEMS for PICs with a focus on scalable PIC platforms. We introduce the fundamental MEMS actuation principles for PICs, review the state of the art quantitatively by comparing reported devices, and provide guidelines for future development.

In Chapter I, we introduce and compare the available tuning mechanisms for PICs (Chapter I.A.), with a focus on MEMS tuning methods (Chapter I.B.). Then, we review the optical functions enabled by MEMS (Chapter I.C.) and compare MEMS actuation mechanisms in terms of their compatibility with PICs (Chapter I.D.). Chapter II presents the current state of the art in MEMS tunable PIC components, with a focus on phase shifters (Chapter II.A), variable couplers (Chapter II.B), switches (Chapter II.C), and beam steering and others (Chapter II.D). Chapter III provides a performance comparison of the

Manuscript received June 15, 2019. This project has received funding from the European Union's Horizon 2020 research and innovation programme under grant agreement No. 780283 (MORPHIC - www.h2020morphic.eu). N. Quack acknowledges funding from the Swiss National Science Foundation (SNSF) under grant No. 157566, and H. Sattari from the Hasler Foundation under grant No. 17008.

C. Errando-Herranz (carloseh@kth.se), P. Edinger (edinger@kth.se), and K. B. Gylfason (gylfason@kth.se) are with KTH Royal Institute of Technology SE-100 44 Stockholm, Sweden.
A. Y. Takabayashi (alain.takabayashi@epfl.ch), H. Sattari (hamed.sattari@epfl.ch) and N. Quack (niels.quack@epfl.ch) are with École Polytechnique Fédérale de Lausanne (EPFL), CH-1015 Lausanne, Switzerland.

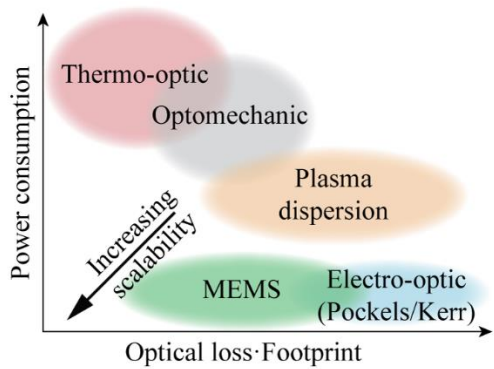

Figure 1. Semi-quantitative comparison between available methods for tuning PICs from a scalability perspective.

TABLE I
COMPARISON OF TUNING MECHANISMS FOR PICs

|  | Power cons. | Optical loss | Foot-print | Speed | Limitations |
| --- | --- | --- | --- | --- | --- |
| Thermo-optic | -- | ++ | + | + | Thermo-optic coefficient |
| Plasma dispersion | + | -- | ++ | ++ | Semiconductors |
| Electro-optic (Pockels/Kerr) | ++ | + | -- | ++ | Crystal symmetry |
| Opto-mechanic | -- | + | + | + | Suspended and optical power dependent |
| MEMS | ++ | + | + | + | Suspended |

++ EXCELLENT    + MEDIUM    -- LIMITED

state of the art of MEMS tunable components, with a focus on large-scale integration in PIC platforms, and outlines the integration prospects for MEMS in PICs. While a special focus is laid on Silicon Photonics, the MEMS concepts apply to the various PIC platforms.

*A. Tuning Mechanisms in PICs*

PIC tuning methods modify waveguide properties by leveraging physical effects such as temperature dependence of refractive index, plasma dispersion in semiconductors, Pockels or Kerr effects for materials with a certain crystal symmetry, and displacement generated by optical gradient forces (optomechanics) or MEMS. Figure 1 shows a visual comparison between available tuning methods in terms of their suitability for application in large-scale PICs.

The choice of tuning method depends on the material platform and the targeted application. For example, for very large-scale PICs working at telecommunication wavelengths, the high refractive index contrast of the silicon-on-insulator (SOI) platform provides a clear advantage in terms of footprint, and offers a wide range of tuning methods, including thermo-optic (due to the large thermo-optic coefficient of silicon [26]), plasma dispersion (silicon is a semiconductor), optomechanics, and MEMS. The other mentioned tuning mechanisms are either not available or negligible, such as the Pockels effect (centro-symmetry of the lattice), and the Kerr effect (very low efficiency) [27]. Silicon nitride waveguides feature less efficient thermo-optic tuning (10 times weaker than in silicon [28]), and lack plasma dispersion mechanisms (SiN is an insulator) and Pockels effect (amorphous structure). III-V compound semiconductors feature plasma dispersion and Pockels tuning but show limited scalability due to their low refractive index contrast and limitations on wafer size [29]. Other than the material platform, key parameters for choosing PIC tuning methods are optical loss, footprint, power consumption, and tuning speed; and each tuning method features inherent trade-offs between these parameters, which are summarized in TABLE I.

Thermo-optic tuning uses the temperature dependence of the refractive index of most materials to tune waveguide properties. It is the most common tuning method for Si and SiN photonics, and features low optical loss and generally small footprint, at the expense of large power consumption, limited speed, and a footprint efficiency limited by the thermo-optic coefficient of the waveguide materials and the thermal cross-talk between adjacent devices. Plasma dispersion tuning relies on the change in refractive index caused by carrier injection or depletion in a material with mobile charge carriers (i.e. semiconductors) [27] and provides small footprint and high speed refractive index tuning, but typically also introduces significant optical loss. The Pockels effect in non-centrosymmetric crystals provides a low-power, low-loss, and high-speed path to refractive index tuning, but requires long propagation lengths (large footprint) due to the low electro-optic coefficients of known waveguide materials. Similarly, the Kerr effect requires prohibitively long devices or exotic materials. Optomechanical tuning relies on optical pumping to generate on-chip optical forces that change the waveguiding properties [25], [30]. This tuning mechanism provides medium/low optical loss and relatively small footprints, with actuation speed limited by inertia of the moving device to the mechanical resonance frequency. A limitation of this method is that, using optical power as the tuning mechanism limits its applications in general-purpose PICs, since, in most cases, they are required to maintain their function independent of the optical power.

In contrast, MEMS photonics rely on a change in optical properties of a waveguide by electromechanical actuation. The low-power and low optical loss of MEMS actuators makes them excellent for large-scale PIC tuning. Although their speed is limited by mechanical resonance frequencies, their working principle is not limited by waveguide material platform, making them widely applicable. Moreover, the design freedom arising from electrically driven mechanical movement, significantly different from the change in refractive index caused by other tuning methods, enables new applications.

*B. Fundamentals of mechanical tuning of PICs*

To understand how light in PICs can be mechanically tuned, it is instructive to look at a generic mathematical expression for a guided electromagnetic mode. In PICs, the waveguide geometries vary considerably less along the direction of propagation than they do in the perpendicular cross-section, and thus we can approach the analysis by breaking the circuits up in segments with fixed cross-sections. Assuming a harmonic time dependence at an angular frequency $\omega$ given by the application, and in the absence of material absorption, we can describe the electric (or magnetic) vector field of an electromagnetic wave travelling in the positive $z$ direction along such a waveguide segment as

$$\vec{E}(x,y,z,t) = \vec{E}_0(x,y)e^{-i(kz-\omega t)} \qquad (1)$$

where $k = 2\pi n_e/\lambda$ is the wave number, $n_e$ is the effective index of the mode, $\lambda$ is the free space wavelength, and $t$ the time. This expression indicates that the mode can be described by a product of a cross-sectional field profile factor $\vec{E}_0(x,y)$ that is maintained during propagation, and a phase factor $e^{-i(kz-\omega t)}$. It is possible to mechanically influence both factors:

1) By applying compressive or tensile stress to the waveguide layers, the materials can be strained, which can modify both the material absorption and refraction [31].

   a) The introduction of material absorption would cause an exponential decay of the mode amplitude with $z$.

   b) A change in waveguide material refractive index would change $n_e$ of the mode, yielding a change in phase shift per step along $z$. The field distribution in the $xy$ plane is also affected.

2) A physical lengthening or shortening of the guide segment along $z$ would modify the phase factor. If this change is achieved by stressing the guide, it would also result in material strain (see 1 above).

3) A displacement of a slab of material within the field of the mode in the waveguide cross-section can yield two effects:

   a) If the moving material slab supports a guided mode of its own that is well phase matched (i.e. has similar $n_e$), light couples between the two, causing a change in field profile and phase of both modes.

   b) If the moving material slab does not support a well-matched mode of its own, the motion only changes the field profile and phase of the waveguide mode. This could be introduction of loss, e.g. by material absorption, radiation or scattering by sidewall roughness, or of a phase shift by adding or removing refractive material.

In practice, all applications require a certain bandwidth, i.e. the waveguides need to operate over a span of angular frequencies, and thus we need to consider that both $\vec{E}_0$ and $n_e$ exhibit frequency dependence (i.e. dispersion). Dispersion can be tuned by effects 1 and 3 above, but 3 is by far the most potent. The strong tuning of mode dispersion by effect 3 is specific to mechanical tuning and differentiates it from most other forms of PIC tuning [32].

By application of the waveguide manipulations described above, several optical functions can be achieved.

*C. Optical functions enabled by MEMS tuning*

Fundamental components of a PIC include linear and nonlinear optical devices, optical sources, and photodetectors, and MEMS can contribute to the function of many of these.

Linear optical devices greatly outnumber other components in PICs, and thus their performance is of central importance. Linear devices include power splitters, filters, delay lines, lenses, mirrors, phase shifters, modulators, and coupling structures for off-chip interfacing. An arbitrary linear optical system can be built using a large enough array of fundamental building blocks. Each building block features two inputs and two outputs and two degrees of freedom to tune the relative power and the phase shift between the two outputs. As phase shifters can be used to build tunable power splitters [33], only phase shifters and passive 2×2 power splitters are strictly necessary. However, specific higher function blocks, such as switching, power splitting, or filtering are often used to reduce footprint and increase optical performance.

As a central component of linear PICs, the main requirement of a phase shifter is that it covers the complete $2\pi$ phase space with minimal optical loss. In general, the phase ϕ gained by an electromagnetic wave at a wavelength λ, travelling in a waveguide mode with effective refractive index $n_e$, over a length L, is

$$\phi = \frac{n_e L}{\lambda} \qquad (2)$$

A phase shift can then be achieved by varying $n_e$ or $L$. Fig. 2a shows a schematic of MEMS actuation approaches to tune $n_e$ and $L$ using displacement or strain.

Most PIC tunable power splitters, based on directional couplers, work by changing the mode interference between two

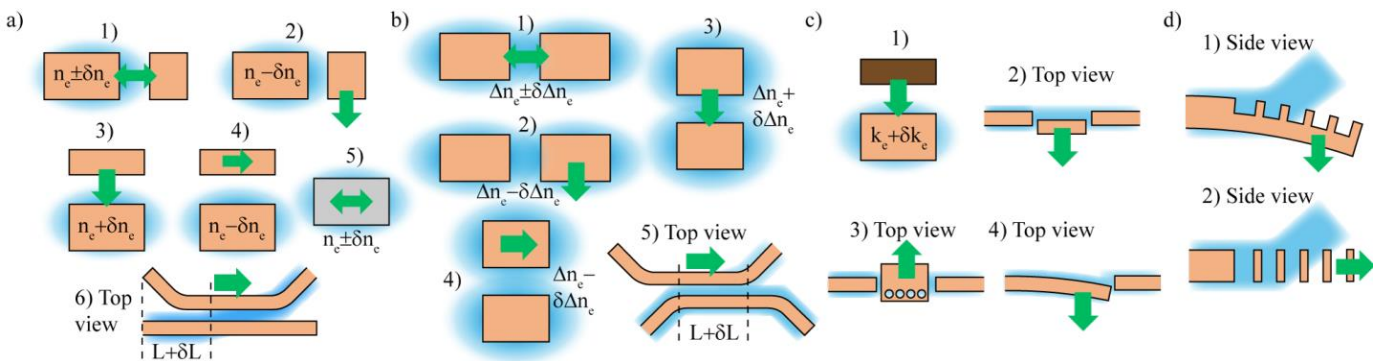

Figure 2. Schematics of MEMS tuning concepts. a) Methods for phase shifting include effective mode index change using waveguide loading with a1) in-plane, a2) single-layer out-of-plane, and a3) double-layer out-of-plane slabs, a4) double-layer in-plane; a5) strain tuning in a piezoelectric material; and a6) effective waveguide lengthening using a movable directional coupler. b) Tunable coupling approaches include b1) in-plane, b2) single-layer out-of-plane, b3) dual-layer out-of-plane and b4) in-plane, and b5) longitudinal tuning. c) Switching methods include c1) displacement of an absorbing material; breaking the waveguide continuity using b2) displacement of a waveguide section, or b3) introduction of a photonic crystal, and b4) bending. d) Tunable grating coupler methods include d1) bending of a grating, and d2) changing its period using in-plane actuation. Unless explicitly mentioned, all figures represent waveguide cross-sections.

adjacent mode-matched waveguides, with the power ratio between the two outputs $P_1$ and $P_2$ defined by

$$\frac{P_2}{P_1} = \sin\left(\frac{\pi L \Delta n_e}{\lambda}\right)^2 \quad (3)$$

with $L$ the coupling length, and $\Delta n_e = n_{e,1} - n_{e,2}$ the difference between the effective mode indexes of the two interfering supermodes. Mechanical tuning can thus be achieved by varying the coupling length $L$ or $\Delta n_e$ by displacement (see Fig. 2b).

Switching a transmitted signal on and off is central to many PIC applications. In conventional PICs, this is achieved by either introducing absorption (using for example plasma dispersion) or using a phase shifter in an interferometric or cavity configuration. However, the displacement enabled by MEMS naturally enables efficient switching. Since there are many approaches to switching, we will only conceptually describe them in this section. MEMS switching can be achieved 1) by using a MEMS tunable directional coupler or a phase shifter in an interferometric/cavity configuration, 2) by changing the waveguide absorption by introducing a lossy medium, 3) by breaking a waveguide continuity by displacing a part of it or 4) by exchanging it for a photonic crystal reflector, and 5) by changing the wave vector direction using bending. Schematics can be found in Fig. 2c.

Tuning of coupling structures to interface PICs with off-chip devices has so far focused on grating couplers. The basic function of a grating coupler can be defined by wave interference, and assuming an in-plane waveguide mode with index $n_e$, leads to

$$\Lambda n_e - \Lambda n_{\text{clad}} \sin\theta = m\lambda \quad (4)$$

with $m = 0, \pm 1, \pm 2 ...$ the diffraction order, $\Lambda$ the grating period, $n_{\text{clad}}$ the mode index in the cladding, $\theta$ the outcoupling angle with respect to the normal, and $\lambda$ the wavelength. Fig. 2d shows methods for MEMS tuning of grating couplers by using displacement to change $\theta$ and $\Lambda$.

Other important functions in PICs are sources, memories, photodetectors, and nonlinear optics. Although these are not the focus of this review, it is worth mentioning that MEMS has been used to control some of these functions, and a brief review can be found in section II.E.

*D. MEMS Actuation Principles for PICs*

The displacements required for MEMS tuning of PIC components lie in the range of tens of nanometers to tens of micrometers, and several types of MEMS actuators, based on different physical principles, fulfill this requirement. The most common MEMS actuation principles are electrostatic, electrothermal, piezoelectric, and magnetic actuation. Here, we provide an overview of the main operating principles suitable for PICs, summarized in Table II. Other known MEMS actuation principles such as pneumatic, shape-memory alloys, electro-active polymers, scratch-drive, phase-change, pyrotechnical, chemical, and biological, are not considered here, due to their lack of integration compatibility with PICs. For a more comprehensive and design-oriented perspective on MEMS actuation, we refer to the review by Bell et al. [34].

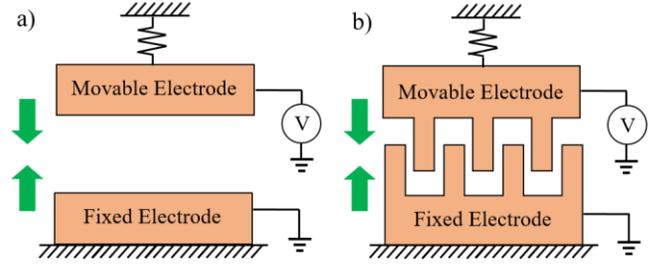

Figure 3. A schematic illustration of a) top/side view of parallel plate and b) top view of comb-drive electrostatic MEMS actuators.

Electrostatic actuators are the most common MEMS actuators. For displacement, they leverage the attractive force between charged plates in a capacitor featuring at least one movable plate. A mechanical spring force counterbalances this attractive force, and results in a stable plate displacement which depends on the potential difference. However, beyond the equilibrium point, the system becomes unstable and the movable plate collapses in a phenomenon known as pull-in. The two most common capacitor geometries used are parallel plate and comb drive arrangements. A similar actuation mechanism, recently explored for PICs, is based on the movement generated by the polarization of a dielectric under an external gradient electrical field. This enables actuation of non-conductive waveguides, which can be significant for material platforms such as SiN.

A schematic visualization of the two, and an illustration of their operation, is given in Fig. 3.

In terms of compatibility with PICs, electrostatic actuation is attractive due to its simplicity. It only requires electrical conductivity in the structural layer and a suitable mechanical suspension. This is particularly true for semiconducting materials such as silicon, in which the actuation mechanism can be integrated within the waveguide layer itself. A limitation

TABLE II
QUALITATIVE ASSESSMENT OF THE MAIN MEMS ACTUATION PRINCIPLES FOR PHOTONIC INTEGRATED CIRCUITS

| Actuation Principle | Photonics Compatibility | Footprint | Response Time | Stroke | Power Consumption | Complexity / Cost | Maximum Displacement | Displacement Resolution |
|---|---|---|---|---|---|---|---|---|
| Electrostatic | ++ | + | ++ | + | ++ | ++ | + / ++ [1] | ++ |
| Thermal | ++ | ++ | + | ++ | -- | ++ | ++ | -- |
| Piezoelectric | + | + | ++ | -- | ++ | + | + | ++ |
| Magnetic | -- | -- | ++ | ++ | -- | + | ++ | ++ |

++ EXCELLENT     + MEDIUM     -- LIMITED

[1] *Parallel Plate / Comb Drive*

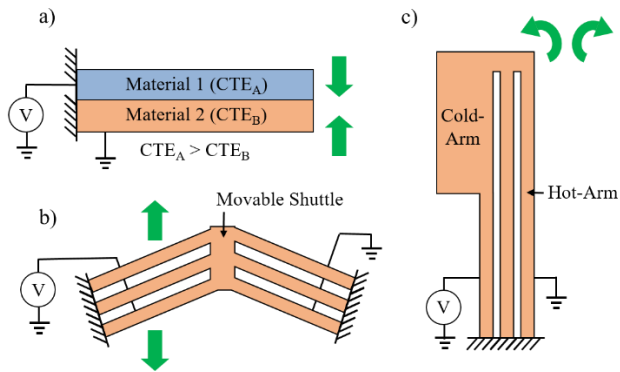

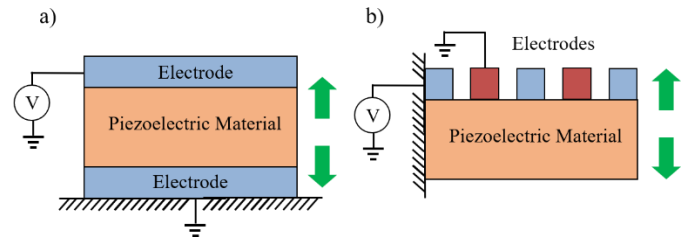

Figure 4. A schematic illustration of typical electrothermal actuators: a) side view of a thermal bimorph actuator, b) top view of a V-shape (or chevron) actuator, and c) a hot-arm/cold-arm actuator.

Figure 5. Applied electric field generates strain in the piezoelectric material, which modulates its length, so the overall structure expands/contracts or bends upwards/downwards: a) side view of a sandwiched electrode configuration, b) and coplanar electrodes.

arising from the use of the waveguide layer as an electrode, particularly in the case of comb drives, is the limited thickness of single mode guides that precludes the use of high-aspect ratio structures that would be advantageous for large force generation. In order to increase the force and stroke, multiple comb-drive sub-units can be connected in parallel to a single device, which, on the other hand, also increases footprint. Response times lie in the microsecond regime and power consumption is minimal due to the absence of current flow under static conditions.

Electrothermal actuators are used for their variety of simple and compact designs and for their large displacements and force output, which allows them to achieve larger stroke than electrostatic actuators. These devices rely on joule heating generated by current flowing through part of the structure, which results in displacement via thermal expansion. There exist two prevalent groups of electrothermal actuators, those based on materials with different coefficients of thermal expansion (CTE) and those using differences in electrical resistance achieved by geometric design. In the first case, the materials form a thermal bimorph, so that when the temperature increases, the difference in expansion between the materials results in displacement. The second type includes the hot-arm/cold-arm and chevron/V-type actuators, which employ geometry to locally modify resistance and, consequently, the amount of thermal expansion. The hot-arm is kept long and narrow (increased resistivity), while the cold-arm is made short and wide (reduced resistivity), so that, under actuation, the structure rotates in the direction of the cold-arm. For the chevron or V-type actuator, linear displacement is achieved by using a V-shape geometry where symmetric thermal expansion causes the structure to move in the direction of the vertex. Visual depictions of these electrothermal actuators are provided in Fig. 4.

Heat exchange becomes very efficient in small structures (due to 1/d dependence, where d is the device dimension), which allows for faster cooling and, consequently, faster switching. However, high-density integration of such actuators is limited by thermal interaction with the PIC and loss of efficiency due to higher chip temperatures. Additionally, because of constant current flow through the device, there is significant power consumption during static operation, making this type of actuation less ideal for low-power applications.

Piezoelectric actuators make use of the piezoelectric effect, in which an applied potential across the piezoelectric material generates strain. These actuators can be made small like electrothermal actuators and responsive like electrostatic actuators but lack the ability to generate large strokes. The need for special piezoelectric materials (crystalline materials without inversion symmetry) and complex fabrication also makes their integration with silicon photonics challenging. Additionally, the generated strain is typically small and must be combined with an elaborate geometry to achieve strokes comparable to the other classes of actuators. Larger strains can be achieved with high strain piezoelectrics and polymers but usually at the cost of reduced output force [34]. Typical geometries consist of a slab of piezoelectric material sandwiched between two electrodes or interdigitated electrodes on top of the piezoelectric. In the first case, an electric field applied between the top and bottom electrodes, generates a negative or positive strain in the horizontal direction of the piezoelectric material, causing it to contract or expand. In the case where the electrodes are interdigitated on top of the piezoelectric, the structure also experiences strain in the horizontal direction. However, because the fixed boundary condition is parallel to the structure, it bends downwards or upwards. In both cases, if the magnitude of the electric field becomes too large (i.e., for large applied voltages and thin piezoelectric layers), the strength of the piezoelectric effect deteriorates. Two common configurations for piezoelectric actuators are shown in Fig. 5.

As is the case with electrostatic actuators, the electric field, not the current flow, drives piezoelectric actuation. Consequently, the power consumption of such devices is small, and the actuation speed can be fast. Additionally, because the functional geometry scales in the out-of-plane direction (i.e., thicker piezoelectric layers generate larger displacements), piezoelectric devices can be made quite compact, in principle. Bulk materials, however, are difficult to integrate with standard MEMS processes and the more compatible thin films do not provide the desired displacement; moreover, piezoelectric actuators typically present a strong hysteresis in the actuation curve, presenting challenges for precision position control. Therefore, additional process development is required before piezoelectric actuators find widespread integration in PICs.

Magnetic actuation is not commonly applied in PICs, but like piezoelectric actuation, offers some distinct advantages at the cost of increased integration complexity. The most common magnetic actuator uses an externally applied magnetic field to attract a patterned permalloy structure or an embedded permanent magnet attached to the movable section [35], [36]. This geometry enables bidirectional motion from the attractive/repulsive electromagnetic force when current is applied (see Fig. 6).

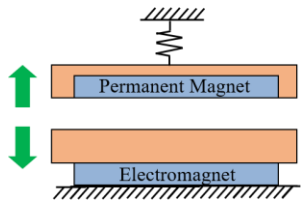

Figure 6. A schematic illustration of a magnetic actuator where the permanent magnets are embedded in the device layer.

Magnetic actuators can generate forces up to hundreds micro-Newtons and achieve displacements in the millimeter range with high resolution [34]. In addition, magnetic actuation provides excellent linearity, which is an advantage for precise feedback control. However, the main disadvantages limiting their use in MEMS is the difficulty of integrating high performance magnetic materials and the large currents required to generate electromagnetic forces.

We note that, in recent years, integration of magnetic materials with photonic waveguides has attracted significant attention due to their applications in Faraday rotation and optical non-reciprocity [37]. These platforms may lead to future demonstrations of magnetic MEMS actuators for photonics.

*E. Latching for MEMS in PICs*

The reduction in PIC power consumption enabled by MEMS can be pushed further by using latching mechanisms. Latching enables non-volatile configuration states where power is only consumed during state transitions. Here, we differentiate between two types of latching mechanisms: 1) latches using additional actuators and 2) latches based on geometrical bistability.

The first type makes use of an additional actuator moving perpendicularly to the primary displacement axis [38]. Both actuators include complementary hooks, such that in the latched state, they come in contact with one another. With such a geometry, the latching actuator can move to "unlatch" the primary actuator, and return after the primary actuator has been displaced, preventing it from following suit and effectively latching it in place (see Fig. 7a). By including multiple hooks, several discrete positions can be addressed, including latching resolution below fabrication resolution by exploiting the Vernier effect [39].

The second type is the bistable latching approach, which employs a beam, or network of beams whose geometry has been selected based on the two stable states existing in a precompressed spring [40]. The structure is fabricated and released in one of the two stable states, but if the applied force by the actuation mechanism exceeds the critical bending force of the beam, the entire device "snaps" to the second stable state (Fig. 7b). The pre-compression can be achieved in-plane by suitable design of the geometry or out-of-plane using compressive stress [41]. In principle, stiction forces can also be used for nonvolatile latching, even though stiction is more difficult to control and thus usually avoided. Push-pull actuators are often used to transition between states in bistable latches and to release structures held in place by stiction [42].

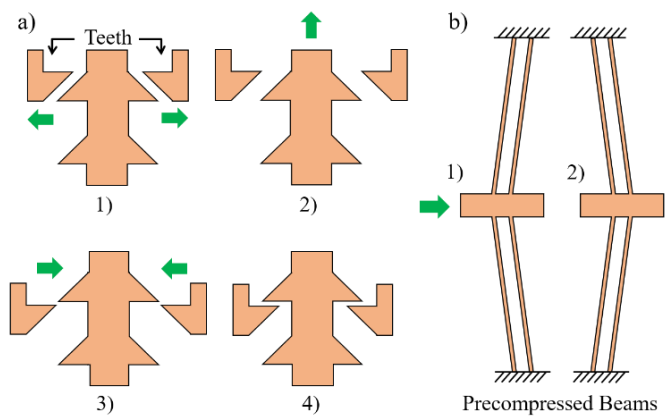

Figure 7. Latching mechanisms for photonic MEMS devices. Top views of a) orthogonal latching dents b) and bistable beams.

## II. REVIEW OF MEMS IN PICs

*A. Phase shifters*

Here we review the MEMS phase shifters described in the literature, including those reported as part of a larger device, e.g. tunable filters or interferometers. Fig. 8 highlights examples from the literature.

A phase shifter on a gold-coated SiN platform using in-plane displacement and parallel-plate actuation, following the concept in Fig. 2a1 was reported in [43]. A similar platform was adapted for out-of-plane displacement of a slab above the waveguide, as in Fig. 2a3, and used both parallel plate and gradient electrical field force actuation [44], including an in-depth analysis of optical loss [45]. In silicon, the concept in Fig. 2a1 was adapted to achieve very efficient in-plane parallel plate actuation by means of a slot waveguide (Fig. 8b) [46], [47]. Comb-drive actuators, providing increased displacement ranges, were used in [48] to increase the magnitude of the phase shift.

Being a fundamental building block, phase shifters, although central to many devices, are often not explicitly reported. A common use of phase shifters is in tunable ring resonators. Kauppinen et al. used out-of-plane displacement of a SiN cantilever on top of SOI waveguides (see schematic in Fig. 2a3) to tune their effective index [49], [50]. Errando-Herranz et al. also used out-of-plane displacement on SOI, but released the ring itself, achieving large phase shifts with a simple fabrication (schematic in Fig. 2a2, SEM in Fig. 8c) [51]–[53]. In-plane tunable ring resonators have also been reported on the SOI platform, relying on suspended directional couplers and comb-drive actuators using the concept in Fig. 2a6, to increase actuation range (Fig. 8d) [54]–[56]. Piezoelectric actuation was utilized to tune the optical path length in SiN resonators [57], [58]. Additionally, strain has been used to tune the effective index of waveguides (Fig. 2a5), and was extensively studied in [59]. This effect was used to phase shift ring resonator waveguides in [60] using a thin film piezoelectric, and in [61] using a bulk piezoelectric substrate. Several phase shifter approaches have been reported as switches. For example, an out-of-plane absorbing membrane was used to tune both real and imaginary parts of the effective index in SiN waveguides, in a combination of the concepts in Figs. 2a3 and 2c1 [62]. Efficient tuning was achieved by plasmonic field enhancement in a compact ring resonator in [63]. In [64], two InP phase shifters using the working principle in Fig. 2a3 were used in an interferometer. MEMS phase shifting in photonic crystal cavities using in-plane actuation has also been reported [42].

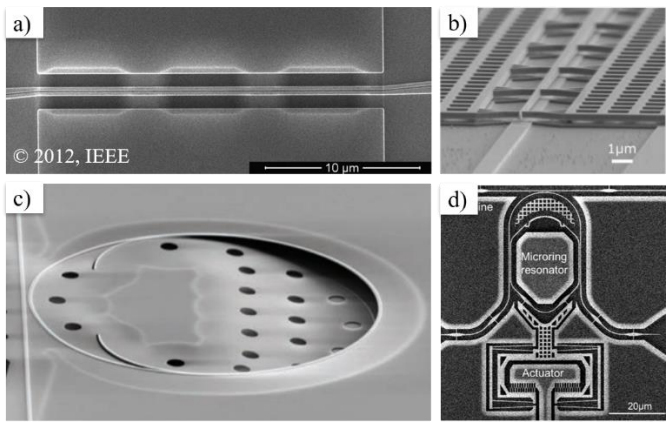

Figure 8. SEM captures of photonic MEMS phase shifters. a) In-plane parallel plate actuation for slot mode tuning [47]. b) Out-of-plane phase shifting with InP [64]. c) Ring resonator filter with out-of-plane SOI actuator [53], and d) In-plane comb-drive actuator for tunable ring resonator length, reproduced from [56] with the permission of AIP Publishing.

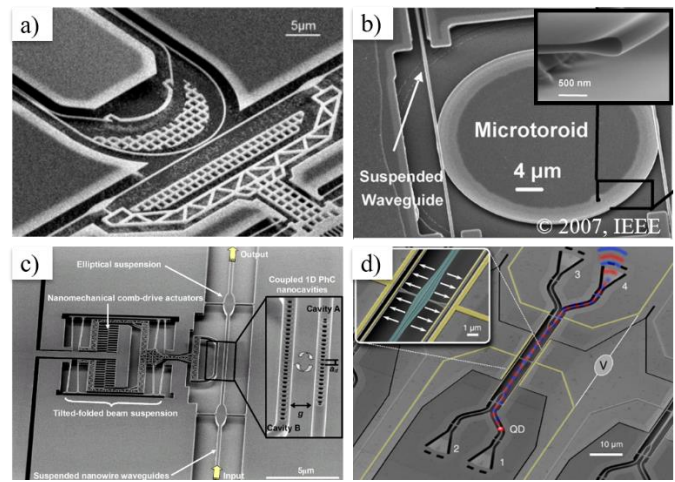

Figure 9. SEM examples of tunable coupling with MEMS. a) Tunable coupling to ring resonator with comb-drive actuation [55]. b) Out-of-plane variable coupler with high-Q microtoroidal resonator [76]. c) In-plane tuning of a dual PhC cavity [78]. d) Single-photon routing with in-plane tunable directional coupler [67]. a), c), d) reprinted with permission from [55], [78], [67] © The Optical Society.

*B. Couplers*

Here we review devices that achieve analog control of power coupling. Figure 9 shows examples of these devices using different electrostatic configurations.

Tunable power coupling is key for efficient PICs, and multiple papers have reported full transduction ranges with directional couplers. A few devices based on III-V platforms have been reported, such as combining InP with the in-plane MEMS tuning concept in Fig. 2b1 [65], or GaAs with out-of-plane (Fig. 2b3) [66], and in-plane [67] parallel-plate actuation. With a similar approach in silicon, tunable coupling was reported using two compact comb drive actuators [68], [69], a simpler single-actuator device [70], and a more stable three actuator system [71]. Tunable couplers using suspended out-of-plane actuation and a single device layer (Fig. 2b3) were also demonstrated [72], [73].

Tunable couplers are widely used to compensate fabrication variations in ring resonators. The tunable ring resonator from Ikeda et al. includes a comb-drive to control the coupling between bus waveguide and ring [55]. Lee et al. used in-plane parallel plate actuation of waveguides to tune light coupling into a disk resonator [74], and out-of-plane displacement for a disk [75] and for a microtoroidal resonator [76]. In a more complex system, Li et al. reported tunable coupling between two resonators [77]. This concept was extended to photonic crystal cavities using in-plane comb-drive actuators [42], [78].

Tunable couplers have also been used to distribute the optical signal from a given input to a set of several outputs at the same time with a well defined distribution ratio of the optical power, commonly referred to as multicast operation [79].

Variable optical attenuators are a special case of couplers that find many uses in PICs, but they have not yet been demonstrated outside of proposals [80].

*C. Switches*

Here we review the various techniques used to implement volatile and non-volatile switching for scalable PICs. A selection of switching mechanisms is shown in Fig. 10.

Bakke et al. provided an early implementation of 1×2 MEMS PIC switch in which an input waveguide was laterally deflected by a pair of comb drives towards one of two output waveguides, such as in Fig. 2c4 [81]. Another type of switch relies on a segmented geometry with a movable waveguide section (schematic in Fig. 2c2, SEM in Fig. 10b) [82]. In-plane movable waveguides have recently been implemented in VTT's micron-scale silicon photonics platform [83], and switching based on frustrated total internal reflection has been demonstrated [84].

Tunable couplers have been commonly used as switches, and for this purpose, Chatterjee et al. demonstrated in-plane perturbation of a static directional coupler using comb-drive actuation [85]. Out-of-plane variants using absorbing metals are also possible, and have been demonstrated using aluminum [62], and gold [86]. Takahashi et al. reported a comb-drive actuated in-plane ring resonator acting as a switch between two bus waveguides [69]. By using the tunable coupler concept in Fig. 2b1, an 8×8 switch matrix was demonstrated [71]. The same concept has been applied to switching using tuning of ring resonators [38], [55].

Out-of-plane switches based on directional couplers (concept in Fig. 2b2) have been scaled up to 50×50 [87], and 240×240 matrices [13], [88]–[93]. Furthermore, polarization independence was demonstrated using two waveguide device layers (Fig. 2b3, SEM in Fig. 10a) [94].

Non-volatile switches in PICs are rare, but a few promising devices have been reported. Abe et al. used comb-drive actuators to displace a latching hook and release it once the main actuator was set [38], [95]. Stiction-based latching has not yet been reported as a non-volatile mechanism, but different reports make use of stiction in other ways. The tunable photonic crystal cavity reported by Chew et al. relies on a push-pull comb-drive to unstick devices [42]. A similar setup could be used for intentional stiction-based nonvolatile latching. The Wu group's results on large-scale switch matrixes utilize stiction and parallel-plate pull-in for reliable bistable operation, even though the switch is volatile [89].

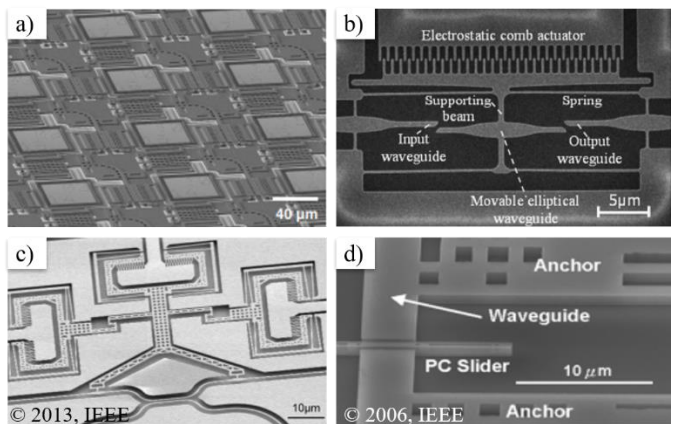

Figure 10. a) Vertical adiabatic coupler switch, reprinted with permission from [89] © The Optical Society. b) Segmented waveguide switch, reproduced from [82] with the permission of AIP Publishing. c) Tunable directional coupler switch with latching [95], d) Photonic crystal switch [96].

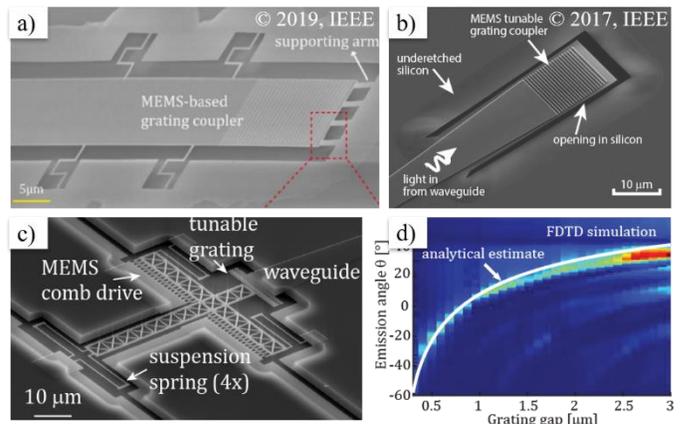

Figure 11. a), b), and c) SEM pictures of reported MEMS tunable grating devices. a) Out-of-plane electrostatically actuated device for spectral tuning [99], and b) chip-to-fiber alignment [98]. c) In-plane comb-drive actuation for beam steering [100], with d) potentially large beam steering, reprinted with permission from [100] © The Optical Society.

Switching has also been demonstrated by using an in-plane actuated switch based on displacing a waveguide/photonic crystal reflector (Fig. 2c3, Fig. 10d) [96]. Switching a photonic crystal on and off was also achieved by using a bimorph cantilever to insert a series of tips aligned into the holes in a photonic crystal waveguide [97].

*D. Grating couplers*

MEMS tuning of grating couplers has only recently been explored, with focus on optical beam steering for optical fiber alignment or sensing, or on tuning of transmission spectra.

By changing the grating angle using out-of-plane electrostatic MEMS tuning of a suspended grating coupler (schematic in Fig. 2d1, SEM in Fig. 11b), Errando-Herranz et al. [98] demonstrated a spatial shift in the coupling to an optical fiber. This approach was adapted by Yu et al. (Fig. 11a) [99] to tune the grating transmission spectra. A second approach to MEMS tuning of grating couplers relies on in-plane actuation using a comb drive to deform a grating coupler shaped like a suspended mechanical spring (Figs. 2d2 and 11c). With such a structure, the period of the grating $\Lambda$ is accessible, potentially resulting the large steering angles, as shown in Fig. 11d) [100].

*E. Integrated sources and nonlinear PICs*

MEMS have been recently leveraged to tune the properties of integrated light sources and nonlinear optic properties of waveguides.

For example, aligning the emission spectra of distinct single-photon sources is central for optical quantum technologies. MEMS-induced strain has been used for this purpose, and groups have reported spectral tuning of quantum dots in a III-V material platform using electrostatic MEMS [101], [102], and in a piezoelectrically-actuated platform combining III-V quantum dots and SiN waveguides [61]. Moreover, electrostatic MEMS in III-V materials have been used to tune the mode volume in a photonic crystal cavity, which in turn tunes the Purcell enhancement, and, in the case of an embedded optical source, its emission rate [103].

For nonlinear PIC applications, MEMS actuation is still relatively unexplored, with proof of concepts using piezoelectric [104] and electrostatic actuation [32] to improve nonlinear optic efficiencies via fine tuning of waveguide birefringence or dispersion.

### III. DISCUSSION

*A. Phase Shifters*

The large number of phase shifters required for large-scale PICs, already reaching the hundreds [105], requires exceptional device performance in terms of optical loss (insertion loss, IL), power consumption, and footprint. In addition, application-dependent figures such as tuning curve linearity, IL variation with actuation, bandwidth, maximum actuation voltage, and induced noise may also be significant.

Since power consumption is minimal for most MEMS actuators, it can be argued that IL is the most relevant figure when discussing the scalability of photonic MEMS circuits. While MEMS phase shifters usually report IL of the same order of magnitude as their thermo-optic counterparts, there are significant differences between designs. In general, the main contributor to losses are transitions (e.g. anchors for waveguide suspension) and waveguide scattering losses due to sidewall roughness. For devices requiring completely suspended waveguides (e.g. single-etch electrostatic actuators), transitions cannot be avoided, and account for IL between 0.1 dB to 1 dB in most devices. Although lower transition losses may be engineered, the large refractive index contrast between silicon and air hinders dramatic improvements, and results in higher waveguide scattering losses compared to oxide-clad devices. Even with state-of-the art silicon PIC foundry platforms, sidewalls typically show rms roughness values of a few nm.

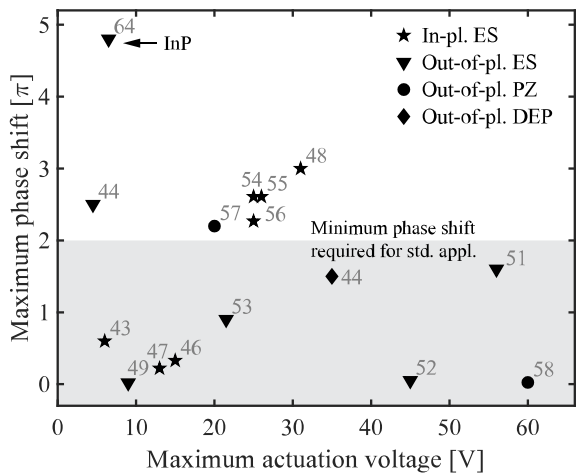

Figure 12. Maximum phase shift versus corresponding voltage for photonic MEMS phase shifters in the literature.

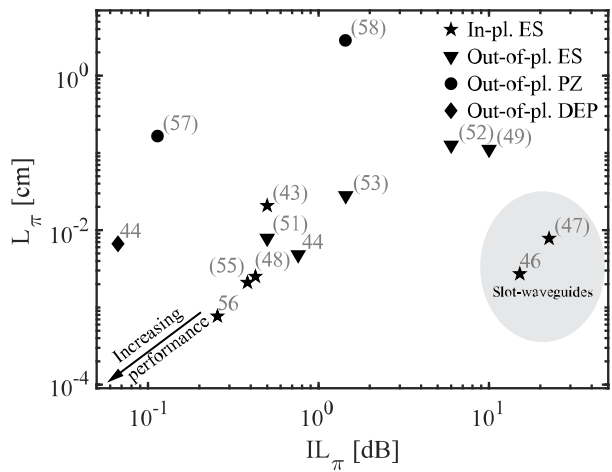

Figure 13. Phase shifter length against IL, both scaled to a π-phase shift. References in parenthesis correspond to estimated values, see Appendix B.

That roughness is the main contributor to standard propagation losses of 1-2 dB/cm in single-mode silicon strip waveguides. Devices that report large phase shifts rely on deconfined waveguides and large evanescent-field interaction, which leads to higher loss due to high power on the sidewalls. This loss dominates when the gap is reduced significantly (i.e. below fabrication resolution). Devices that rely on out-of-plane displacement of a slab on top of a waveguide, on the other hand, benefit from the near-atomically-flat top and bottom surfaces, and usually feature lower roughness-induced losses. However, reported devices using such an architecture have so far relied on beams that are not entirely phase mismatched to the guided mode, effectively becoming a lossy asymmetric directional coupler [44], [45], [64].

Regarding variable loss with actuation, phase shifters that use displacement of beams near waveguides fall into two categories. Devices that rely on reducing a gap below fabrication resolution can see their IL increase with actuation, while devices that increase the original gap can see a decrease of roughness-induced loss. On the other hand, the insertion loss of phase shifters that use in-plane displacement of a stable directional coupler [48] scales with the usually lower waveguide loss, although they may suffer from additional variable IL due to coupling instability. Piezoelectric-based phase shifting, in contrast, does not require deconfined waveguides nor transitions, and as a result, insertion loss can be much lower. The dominant contribution becomes the propagation losses, and even with the largest reported structures, the associated IL remains low [57], [58], [60], [61].

To be able to access all the phase space, a phase shifter must be able to address any shift within $2\pi$. Figure 12 summarizes the maximum phase shift and actuation voltages reported in the literature, and shows that only a few demonstrate a significant maximum phase shift, although, in principle, any of the underperforming devices could be cascaded up to $2\pi$.

From Fig. 12, we can conclude that usual MEMS actuation in PICs requires voltage levels below 100 V. We should note that, in MEMS, the actuation voltage is usually a design choice, which trades off with actuation speed through the designed spring constant. A significant technology-dependent feature is the shape of the actuation curve. Evanescent phase shifting is associated to highly nonlinear actuation curves due to the exponential distribution of the evanescent field [46]. The effect can be compensated for by using gap-increasing actuation and leveraging the quadratic response of electrostatic actuators, yielding a quasi-linear actuation range [53]. Another option for better linearity is to change the propagation length instead, removing the exponential dependence of a gap change. Linear phase shifts can also be achieved using piezoelectrics [58], [60], [61], and quadratic with comb-drive actuators [54], [55].

Device footprint is not specific to phase shifters, but we will highlight a few points here. Shorter interaction lengths are typically required to achieve full power exchange in a coupler, than required for a $2\pi$ shift in a phase shifter. The largest phase shifts with silicon waveguides are achieved with in-plane increase of the propagation length by using suspended directional couplers, which comes at the cost of higher IL due to the couplers loss and sidewall scattering [48]. The InP phase shifters reported in [64] also stand out in Fig. 12, although their IL, which is expected to be significant due to waveguide anchors and coupler asymmetry, was not reported.

Phase shifters can be cascaded and still seen as a single component, while couplers and switches would require more intricate circuit design to be cascaded. As a result, IL and length/footprint values don't have much significance unless scaled by the phase shift. In Fig. 13 we use such scaled figures to compare the different actuation principles in terms of static losses and interaction length. Since measured IL values are missing in many reports, some points were estimated using other reported measurements (see Appendix B). Although instructive, care should be taken when analyzing technological trends from this plot, since not all devices scale in the same way. As an example, the IL of the phase shifter in [48] will have a similar value whether designed for very low phase shifts or large ones, due to the need for the two directional couplers with IL of 0.2 dB each. Even with the conservative analysis required, due to estimates and scaling, some trends clearly stand out. First, most electrostatic devices seem to follow a general rule:

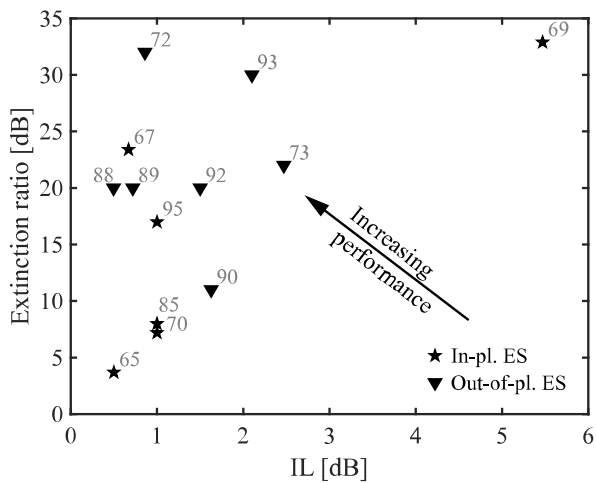
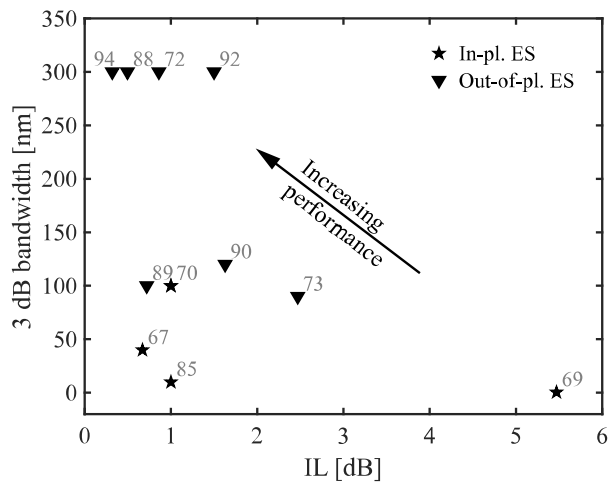

Figure 14. Extinction ratio versus insertion loss for couplers and switches.

Figure 15. 3 dB bandwidth versus insertion loss for couplers and switches.

tuning efficiency (associated to $L_\pi$) scales with IL. Such a result suggests that the optical losses are defined mostly by length, probably dominated by waveguide and anchor loss. Only the slot waveguides stand out among electrostatic phase shifters, due to the higher IL associated to slot modes and mode converters [46], [47]. Finally, when comparing the different electrostatic actuators, it appears that in-plane phase shifters attain better performance. However, both the reported in-plane and out-of-plane actuators actually comprise various tuning mechanisms, and we do not believe the differences to be set in stone. Piezoelectric phase shifters are a promising alternative due to the very low IL attained, although they require about an order of magnitude longer devices. Actuators based on a gradient electric field (dielectrophoresis, DEP) are rare, but show great promise as an alternative to parallel-plate actuation, in terms of $IL_\pi$ (even though it may be due to the lower index contrast of SiN/air and to the flat bottom/top surfaces), and due to the absence of pull-in, effectively increasing possible tuning range and stability.

With the development of large-scale reconfigurable circuits, we expect a significant increase of research on MEMS phase shifters. For proper comparison and valid scalability claims, we suggest that future reports include measurements of IL, variable IL, reflections, and actuation linearity. Additionally, a definition of bandwidth should be introduced, that can then be used for the design of spectral functions in large-scale circuits.

*B. Couplers and Switches*

As is the case for phase-shifters, the number of couplers and switches employed in large-scale PICs is steadily increasing, with the performance of individual devices determining the overall performance of the system. Certain key parameters can be used to assess this performance, which include IL, extinction ratio, bandwidth, switching speed or response time, device footprint, actuation voltage, and power consumption.

Figure 14 provides a graphic comparison of published couplers and switches with regard to extinction ratio and IL. We do not observe a clear trend between in-plane and out-of-plane devices in terms of ER and IL, although out-of-plane devices seem to achieve slightly higher ER in general. Similarly, no clear trend can be observed in terms of IL, and the differences between devices can be ascribed to optical design and fabrication. However, we should note that the couplers based on two waveguide levels exhibit slightly lower IL, most probably due to the smaller top and bottom surface roughness, as compared to the sidewall roughness in single-layer devices. A clear outlier is [69], with a high IL due to lossy anchors.

Figure 15 compares reported IL and 3-dB bandwidth for various couplers and switches. Here, devices that feature adiabatic couplers [72], [88], [92], [94], yield significantly wider bandwidth than directional couplers. This is due to the absence of periodic exchange of power associated with mode interference. In addition, the devices with the lowest reported bandwidth are switches based on cavities.

The actuation voltage for couplers and switches ranges between 5 V for cavity-based devices to the more common 30 to 50 V range for large waveguide structures [77]. Indeed, switches relying on cavities are more sensitive to actuation, and do not require as much displacement as their non-resonant counterparts and can therefore operate at lower voltages, at the cost of a resonance-limited bandwidth.

As next-generation devices move towards the upper left-hand corner of the two plots in Figs. 14 and 15 and become faster and more compact, it is important that all discussed parameters continue to be reported for benchmarking performance. Furthermore, large-scale PICs particularly suffer from back reflection, which can distort the quality of the signal, especially in bidirectional meshes [106]. In order to evaluate couplers and switches according to this metric, reflection should be measured and reported. Currently in-plane and out-of-plane electrostatic devices monopolize the coupler/switch field, but the problem space remains open for new combinations of actuators and optical designs.

*C. Grating couplers*

Quantitative assessment of the performance of tunable grating couplers is highly dependent on the targeted application. Here, we focus on three application areas identified in literature: mode matching, spectral tuning, and free space beam steering.

Mode matching is highly dependent on the devices to be interfaced. So far, the only investigated application has been in fiber-to-chip alignment, with demonstrated 6 μm maximum tuning with 6 V along one direction [98], at the expense of varying transmission efficiency. Future work in this direction can benefit from additional tuning in the perpendicular direction, by e.g. tilting the device sideways in a similar way to MEMS micromirrors [19].

Tuning the transmission spectra of grating couplers allows for compensation of fabrication variation by shifting the optimum wavelength $\Delta\lambda_{Tmax}$ to cover the expected fabrication variation, while minimizing variation in optical bandwidth $\delta\lambda$ and in coupling efficiency. The only presented MEMS tunable grating coupler targeting this application achieves a $\Delta\lambda_{Tmax}$ of 22.8 nm with 12 V actuation [99]. This performance is comparable to power-hungry thermo-optic tuning of grating couplers [107], [108], including MEMS-mediated thermo-optic designs [109], which in general feature $\delta\lambda$ and efficiency variations well below the 6 nm and 0.5 dB reported by the MEMS device.

Free space beam steering features a different set of requirements, such as 2-dimensional steering, large angular tuning $\Delta\theta$, and low angular lobe width $\delta\theta$, with minimum variation with actuation. A sensible figure of merit is their ratio $\Delta\theta/\delta\theta$. Most applications also require a high tuning speed, a stable coupling efficiency, and a minimum optical spectral bandwidth if the device relies on wavelength scanning for steering along one dimension. The device presented in [100] can potentially achieve large $\Delta\theta$ up to 90°. However, experimentally, the device achieved up to 5.6° for 20 V actuation, with angular full-width half-maximum of 9° and 14° along the tuning direction and the perpendicular, i.e. a $\Delta\theta/\delta\theta$ of 0.62. Simulations yielded a variable coupling efficiency from 30 to 40% and a tuning speed up to 200 kHz. As a comparison, last-generation optical phased arrays (OPAs) based on plasma dispersion phase shifters feature $\Delta\theta/\delta\theta$ of 500, with tuning speeds in the order of 30 kHz, at the expense power consumption and footprint that were orders of magnitude larger [105]. Improved performance can potentially be achieved by MEMS actuator design focused on stability and long displacements, and higher speeds through larger spring constants. 2D steering can be achieved by nesting MEMS actuators, or by integrating the devices into a linear OPA. An alternative approach would be to use MEMS phase shifters in an OPA to potentially combine the low-power consumption of MEMS with the optical performance of OPAs.

We can now identify a few general guidelines for reporting and designing MEMS tunable grating couplers. Future reporting of results should preferably include a full set of measurements including wavelength dependence, angular emission using Fourier imaging, coupling efficiency, and reflection. With these measurements, design may focus on improving the previously introduced figures of merit, while minimizing drifts in efficiency or optical bandwidth. This goal can be achieved by careful design of MEMS devices, or by combining MEMS tuning with other tuning methods. This field is still in its infancy, and a wide range of novel ideas that further

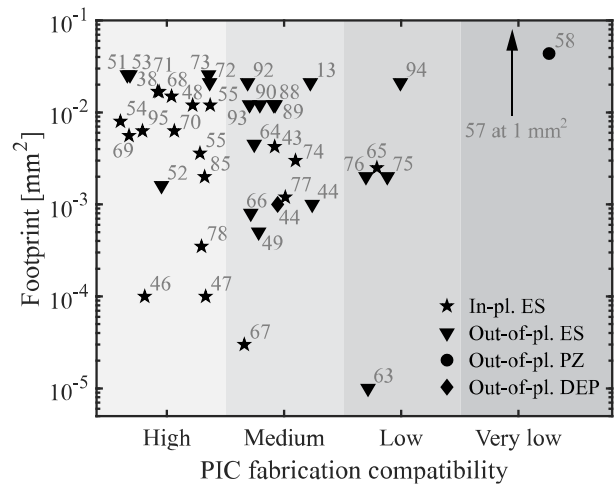

Figure 16. Device footprint versus PIC technology compatibility score. Devices include phase shifters, couplers and switches. More information on the derivation of the compatibility score in Appendix C.

exploit the large geometrical variations enabled by MEMS has yet to be explored.

*D. Perspectives on Large-Scale Integration of MEMS in PICs*

Large-scale photonic circuits would benefit from MEMS actuation due to its low power consumption. PIC platforms (silicon photonics in particular), and MEMS have grown rapidly thanks to the development of processes for the microelectronics industry. As a result, dedicated photonics and MEMS platform have a lot in common, and a photonic MEMS platform is only a step away [110], [111]. The key components for reprogrammable photonic MEMS circuits have already been reported, and large scale switch networks have been achieved with wafer-scale processes [71], [89]. Further improvements on the analog building blocks can be expected in the coming years, with demonstrators of reprogrammable PICs beyond switch networks. Here we discuss the photonic MEMS trade-offs common to all building blocks, and how those trade-offs impact scalability.

Minimizing IL is key for PIC scalability and has been extensively discussed in sections III.A. and III.B. As fabrication processes improve, and foundries become accessible, we can expect standardized MEMS phase shifters and couplers with IL below 0.1 dB. As with microelectronics, footprint is one of the main drivers for chip cost and the footprint of a single chip depends on the aggregate size of individual components. While devices should be kept as small as possible, the number of driving contacts required also contribute to the footprint. Furthermore, a complex interposer and the associated packaging may dominate the cost [112]. The type of interposer depends not only on the number of contacts, but also on their voltage levels, since higher voltages require larger interconnect spacing to minimize crosstalk. This property is a problem in particular for piezoelectric and DEP MEMS actuators, with actuation voltages above 50 V. In contrast, the driving voltage for electrostatic MEMS actuators can be much lower, but sub –10 V devices are usually associated with low spring stiffness (below 0.1 N/m) making them susceptible to noise.

The success of the reported phase shifters, couplers or switches as building blocks for scalable circuit depends heavily on materials and processing technology. The integration of non-standard materials, such as thin film piezoelectrics, into standard platforms will require additional process development. Reported devices show different levels of compatibility with open-access foundries and was assessed on a compatibility scale (more information in Appendix C). Complexity and footprint are the main drivers for cost, and Fig. 16 shows a visualization of how devices with a similar complexity are distributed with regard to footprint. On such a scale, the slot waveguides from Acoleyen et al. clearly stand-out [46], [47], and an improvement on the IL, caused by the lossy slot modes and mode converters, would make them excellent phase shifters. The tunable PhC cavity published by Chew et al. demonstrates that even simple comb-drive actuators can be very competitive in terms of footprint [78]. Although reported piezoelectric actuators do not appear advantageous according to this metric, as discussed in III.A., phase shifters using piezoelectric materials report the lowest IL thanks to their high waveguide confinement, and the absence of scattering transitions. Consequently, special applications where footprint is not a limitation could benefit from the integration of piezoelectric materials in PIC foundries.

Photonic MEMS devices typically have resonant frequencies of 100 kHz to 10 MHz, which limits actuation speed. Out-of-plane electrostatic actuators achieve switching speeds above 2 MHz, while in-plane actuators report speeds in the order of 100 kHz. The switching speed of MEMS is in general linked to their spring constant and mass, which are design parameters, suggesting that tens of MHz speeds are within reach. Moreover, the electric time constant might pose an additional limitation [71]. Therefore, we do not expect MEMS to substitute current high-speed opto-electronics modulation. However, a combination of low-power MEMS reconfiguration with the existing high-speed modulation in photonics foundries is well within reach.

The resolution of photonic MEMS is defined by 1) the precision in the analog driving voltage, or, for photonic ICs, by the number of bits of the Digital to Analog Converter (DAC), and 2) the actuation curve (e.g. phase shift or coupling ratio versus voltage. This discussion is strictly linked to the effect of electrical noise. For example, nonlinear response curves will therefore have actuation-dependent fluctuation that can be detected in outputs like phase shifts [113], and can be very detrimental for certain PIC architectures [106]. Possible solutions are to increase actuation linearity [113] or to use custom, more costly DACs. Another source of noise, which is common in MEMS, is environmental variation in suspended geometries, usually caused by humidity or particles. Such disturbances are typically minimized by hermetic packaging [114], [115], which will likely be used for MEMS-based PICs as well. Thermomechanical noise may also cause fluctuations in small gap spacing; however, at room temperature operation, even with low mechanical quality factor, these perturbations are on the picometer-scale and do not present an issue.

Photonic MEMS devices will be subject to reliability of their actuation mechanism. The reliability of MEMS actuators has been studied extensively in previous reports [116], and adequate strategies to overcome reliability-related limitations in MEMS have been developed [117]. For silicon photonic MEMS, a number of effects might lead to reliability limitations. Any variation in geometric definition of the actuator will lead to a variation in actuation voltage [92]. Such variations can be kept within acceptable levels with modern lithography capabilities. Electrostatic MEMS actuators can further present hysteresis caused by charge trapping in a dielectric located between the plates. Hysteresis can be reduced by avoiding high-k dielectrics between the parallel plates (i.e. by complete removal of the buried oxide in SOI), and by actuation approaches such as polarity reversal. While fatigue failure is not observed in silicon due to the crystalline nature of the material, extreme environments and high levels of mechanical stress exceeding the yield strength will lead to fracture, which can be prevented by adequate design and operating conditions. For MEMS requiring contact, such as latching systems, wear is a significant failure mechanism, and can set strict limits on repeatability and device lifetime.

While low power consumption is the most prominent benefit of MEMS actuation, the minimum power consumption is limited by nonzero leakage currents, thereby requiring a constant voltage supply. Non-volatility using latching enables zero-power photonics and novel functionalities such as one-time reconfiguration to compensate for fabrication variation.

The introduction of on-chip suspended waveguides and MEMS provides new ground for innovation and new applications for PICs. A few early examples are tunable grating couplers [100], dispersion tuning [32], optomechanical components for nonlinear and nonreciprocal optics [25], and Brillouin lasers and amplifiers [118].

A recent application is quantum photonics, with MEMS providing a clear advantage over other tuning methods due to the strict requirements for low operating temperatures and low optical loss, and the possibility of strain tuning of quantum sources [61], [66], [67], which additionally improves qubit stability and coherence in certain systems [119]–[121]. These properties, combined with the need for large number of phase shifters for applications in quantum simulation and computing [15], makes MEMS an excellent tuning method for quantum photonics, and we can expect breakthroughs in quantum PICs enabled by MEMS in the near future.

## IV. Conclusion

In summary, we have described the field of MEMS for photonic integrated circuits. A review of experimentally demonstrated devices is provided, with a special focus on the basic building blocks for large-scale PICs, including phase shifters and couplers/switches. Specifically, device performance and prospects for integration into large-scale PICs are discussed and quantitatively assessed. The insights obtained provide relevant guidelines to design MEMS for advanced PICs, which is of particular value given current efforts aimed at adopting MEMS technology as a standard process module in photonics foundries [122]. The widespread integration of MEMS in PICs will provide new functionality and can be combined with existing advanced photonics modules, such as high-speed photodetectors and modulators. As such, designers will have access to a powerful toolbox of advanced photonics capabilities for low-loss, low-power and high-performance PICs for applications in information and communication technologies, sensors for consumer electronics, LIDAR 3D imaging, biosensing, quantum sensing or quantum information processing.

APPENDIX

*A. Extraction of phase shift*

Tunable ring resonators relying on phase shifting often do not report the maximum phase shift required for their reported resonant wavelength shift $\Delta\lambda$. When the free spectral range (FSR) was stated, we include the paper in the FoM plots in III.A. by extracting the phase shift as $\Delta\phi = \Delta\lambda/\text{FSR}$.

*B. Extraction of insertion loss*

Many reports do not include a measurement of IL. IL is a key figure and, when possible, we estimated it based on the actuation principle and platform used as follows. We did not include propagation loss in the calculation, except for particularly long devices, lossy propagation modes, and devices with only propagation losses as a source of loss. When not discussed, we included the following tapering losses for SOI devices: 0.05 dB/transition for in-plane tapered anchors, when key to the device functionality; and 0.1 dB/transition for changes in the cladding (e.g. transition from suspended to strip waveguide). For tunable ring resonators relying on phase shifts, we calculated the IL as $\text{IL[dB]} = \alpha[\text{dB/cm}] \times C = 2\pi\lambda_0/Q_{\text{int}}.\text{FSR}$, with $C$ the round-trip length, and $Q_{\text{int}}$ the intrinsic quality factor, approximated as the loaded quality factor when not available. If only a small portion of the round-trip is used for phase shifting, we subtract a propagation loss corresponding to the larger passive portion of the resonator.

*C. Photonics compatibility assessment*

In order to compare the performance and cost of components for large scale circuits, we assessed them on a "photonics compatibility" scale, indicating devices requiring only 1-2 steps of post-fabrication processing (*High*); more than 3 simple post-processing steps but still compatible with current open foundries (*Medium*); complex post-processing with standard materials (*Low*); and non-standard materials and processes that are not foundry-compatible (*Very low*). When plotted, we added a random spread to devices with the same grade to improve readability.

*D. Data availability*

The data behind the discussion in Section III, comprising the extracted and calculated key figures from the literature, can be found in the linked dataset [123].

REFERENCES

[1] C. R. Doerr, "Silicon photonic integration in telecommunications," *Front. Phys.*, vol. 3, Aug. 2015.
[2] S. Rumley, D. Nikolova, R. Hendry, Q. Li, D. Calhoun, and K. Bergman, "Silicon Photonics for Exascale Systems," *J. Light. Technol.*, vol. 33, no. 3, pp. 547–562, Feb. 2015.
[3] K. De Vos, I. Bartolozzi, E. Schacht, P. Bienstman, and R. Baets, "Silicon-on-Insulator microring resonator for sensitive and label-free biosensing," *Opt. Express*, vol. 15, no. 12, pp. 7610–7615, 2007.
[4] J. W. Silverstone, D. Bonneau, J. L. O'Brien, and M. G. Thompson, "Silicon Quantum Photonics," *IEEE J. Sel. Top. Quantum Electron.*, vol. 22, no. 6, pp. 390–402, Nov. 2016.
[5] M. Pantouvaki *et al.*, "50Gb/s silicon photonics platform for short-reach optical interconnects," in *Optical Fiber Communication Conference*, 2016, pp. Th4H–4.
[6] E. Timurdogan *et al.*, "AIM Process Design Kit (AIMPDKv2.0): Silicon Photonics Passive and Active Component Libraries on a 300mm Wafer," in *2018 Optical Fiber Communications Conference and Exposition (OFC)*, 2018, pp. 1–3.
[7] K. Giewont *et al.*, "300-mm Monolithic Silicon Photonics Foundry Technology," *IEEE J. Sel. Top. Quantum Electron.*, vol. 25, no. 5, pp. 1–11, Sep. 2019.
[8] P. Muñoz *et al.*, "Silicon Nitride Photonic Integration Platforms for Visible, Near-Infrared and Mid-Infrared Applications," *Sensors*, vol. 17, no. 9, p. 2088, Sep. 2017.
[9] A. E.-J. Lim *et al.*, "Review of Silicon Photonics Foundry Efforts," *IEEE J. Sel. Top. Quantum Electron.*, vol. 20, no. 4, pp. 405–416, Jul. 2014.
[10] "Europractice Homepage." [Online]. Available: http://www.europractice-ic.com/. [Accessed: 12-Jun-2019].
[11] "MOSIS Integrated Circuit Fabrication Service." [Online]. Available: https://www.mosis.com/. [Accessed: 12-Jun-2019].
[12] J. Sun, E. Timurdogan, A. Yaacobi, E. S. Hosseini, and M. R. Watts, "Large-scale nanophotonic phased array," *Nature*, vol. 493, no. 7431, pp. 195–199, Jan. 2013.
[13] T. J. Seok, N. Quack, S. Han, R. S. Muller, and M. C. Wu, "Highly Scalable Digital Silicon Photonic MEMS Switches," *J. Light. Technol.*, vol. 34, no. 2, pp. 365–371, Jan. 2016.
[14] A. Ribeiro, A. Ruocco, L. Vanacker, and W. Bogaerts, "Demonstration of a 4x4-port universal linear circuit," *Optica*, vol. 3, no. 12, pp. 1348–1357, Dec. 2016.
[15] J. Wang *et al.*, "Multidimensional quantum entanglement with large-scale integrated optics," *Science*, vol. 360, no. 6386, pp. 285–291, Apr. 2018.
[16] N. C. Harris *et al.*, "Quantum transport simulations in a programmable nanophotonic processor," *Nat. Photonics*, vol. 11, no. 7, pp. 447–452, Jul. 2017.
[17] Y. Shen *et al.*, "Deep learning with coherent nanophotonic circuits," *Nat. Photonics*, vol. 11, no. 7, pp. 441–446, Jul. 2017.
[18] D. Pérez *et al.*, "Multipurpose silicon photonics signal processor core," *Nat. Commun.*, vol. 8, no. 1, p. 636, Sep. 2017.
[19] M. C. Wu, O. Solgaard, and J. E. Ford, "Optical MEMS for Lightwave Communication," *J. Light. Technol.*, vol. 24, no. 12, pp. 4433–4454, Dec. 2006.
[20] A.-Q. Liu, *Photonic MEMS Devices: Design, Fabrication and Control*. CRC Press, 2018.
[21] F. Chollet, "Devices Based on Co-Integrated MEMS Actuators and Optical Waveguide: A Review," *Micromachines*, vol. 7, no. 2, p. 18, Jan. 2016.
[22] H. Du, F. Chau, and G. Zhou, "Mechanically-Tunable Photonic Devices with On-Chip Integrated MEMS/NEMS Actuators," *Micromachines*, vol. 7, no. 4, p. 69, Apr. 2016.
[23] L. Midolo, A. Schliesser, and A. Fiore, "Nano-opto-electro-mechanical systems," *Nat. Nanotechnol.*, vol. 13, no. 1, p. 11, Jan. 2018.
[24] M. Stepanovsky, "A Comparative Review of MEMS-Based Optical Cross-Connects for All-Optical Networks From the Past to the Present Day," *IEEE Commun. Surv. Tutor.*, vol. 21, no. 3, pp. 2928–2946, thirdquarter 2019.
[25] M. Aspelmeyer, T. J. Kippenberg, and F. Marquardt, "Cavity optomechanics," *Rev. Mod. Phys.*, vol. 86, no. 4, pp. 1391–1452, Dec. 2014.
[26] H. H. Li, "Refractive index of silicon and germanium and its wavelength and temperature derivatives," *J. Phys. Chem. Ref. Data*, vol. 9, no. 3, pp. 561–658, 1980.
[27] R. Soref and B. Bennett, "Electrooptical effects in silicon," *IEEE J. Quantum Electron.*, vol. 23, no. 1, pp. 123–129, Jan. 1987.
[28] R. Amatya, C. W. Holzwarth, H. I. Smith, and R. J. Ram, "Efficient Thermal Tuning for Second-order Silicon Nitride Microring Resonators," in *Photonics in Switching, 2007*, 2007, pp. 149–150.

[29] D. Liang and J. E. Bowers, "Photonic integration: Si or InP substrates?," *Electron. Lett.*, vol. 45, no. 12, pp. 578–581, Jun. 2009.
[30] T. J. Kippenberg and K. J. Vahala, "Cavity Opto-Mechanics," *Opt. Express*, vol. 15, no. 25, pp. 17172–17205, Dec. 2007.
[31] Y. Kim, M. Takenaka, T. Osada, M. Hata, and S. Takagi, "Strain-induced enhancement of plasma dispersion effect and free-carrier absorption in SiGe optical modulators," *Sci. Rep.*, vol. 4, p. 4683, Apr. 2014.
[32] C. Errando-Herranz, P. Edinger, and K. B. Gylfason, "Dynamic dispersion tuning of silicon photonic waveguides by microelectromechanical actuation," in *Lasers and Electro-Optics (CLEO), 2017 Conference on*, 2017, pp. 1–2.
[33] D. A. B. Miller, "Perfect optics with imperfect components," *Optica*, vol. 2, no. 8, pp. 747–750, Aug. 2015.
[34] D. J. Bell, T. J. Lu, N. A. Fleck, and S. M. Spearing, "MEMS actuators and sensors: observations on their performance and selection for purpose," *J. Micromechanics Microengineering*, vol. 15, no. 7, pp. S153–S164, Jul. 2005.
[35] H. J. Cho and C. H. Ahn, "A bidirectional magnetic microactuator using electroplated permanent magnet arrays," *J. Microelectromechanical Syst.*, vol. 11, no. 1, pp. 78–84, Feb. 2002.
[36] M. Khoo and C. Liu, "Micro magnetic silicone elastomer membrane actuator," *Sens. Actuators Phys.*, vol. 89, no. 3, pp. 259–266, Apr. 2001.
[37] B. J. H. Stadler and T. Mizumoto, "Integrated Magneto-Optical Materials and Isolators: A Review," *IEEE Photonics J.*, vol. 6, no. 1, pp. 1–15, Feb. 2014.
[38] S. Abe and K. Hane, "A silicon microring resonator with a nanolatch mechanism," *Microsyst. Technol.*, vol. 21, no. 9, pp. 2019–2024, Sep. 2015.
[39] A. Unamuno and D. Uttamchandani, "MEMS variable optical attenuator with vernier latching mechanism," *IEEE Photonics Technol. Lett.*, vol. 18, no. 1, pp. 88–90, Jan. 2006.
[40] H. Liu and F. Chollet, "Moving Polymer Waveguides and Latching Actuator for 2 $ \times$ 2 MEMS Optical Switch," *J. Microelectromechanical Syst.*, vol. 18, no. 3, pp. 715–724, Jun. 2009.
[41] H. Sattari, A. Toros, T. Graziosi, and N. Quack, "Bistable silicon photonic MEMS switches," in *MOEMS and Miniaturized Systems XVIII*, San Francisco, United States, 2019, p. 13.
[42] X. Chew *et al.*, "An in-plane nano-mechanics approach to achieve reversible resonance control of photonic crystal nanocavities," *Opt. Express*, vol. 18, no. 21, pp. 22232–22244, Oct. 2010.
[43] M. Poot and H. X. Tang, "Broadband nanoelectromechanical phase shifting of light on a chip," *Appl. Phys. Lett.*, vol. 104, no. 6, p. 061101, Feb. 2014.
[44] M. W. Pruessner, D. Park, T. H. Stievater, D. A. Kozak, and W. S. Rabinovich, "Broadband opto-electro-mechanical effective refractive index tuning on a chip," *Opt. Express*, vol. 24, no. 13, pp. 13917–13930, Jun. 2016.
[45] M. W. Pruessner, D. Park, B. J. Roxworthy, T. H. Stievater, N. F. Tyndall, and W. S. Rabinovich, "Loss reduction in electro-mechanically-tunable microring cavities," *Opt. Lett.*, p. 6, 2019.
[46] K. Van Acoleyen, J. Roels, T. Claes, D. Van Thourhout, and R. Baets, "NEMS-based optical phase modulator fabricated on Silicon-On-Insulator," in *Group IV Photonics (GFP), 2011 8th IEEE International Conference on*, 2011, pp. 371–373.
[47] K. V. Acoleyen, J. Roels, P. Mechet, T. Claes, D. V. Thourhout, and R. Baets, "Ultracompact Phase Modulator Based on a Cascade of NEMS-Operated Slot Waveguides Fabricated in Silicon-on-Insulator," *IEEE Photonics J.*, vol. 4, no. 3, pp. 779–788, Jun. 2012.
[48] T. Ikeda, K. Takahashi, Y. Kanamori, and K. Hane, "Phase-shifter using submicron silicon waveguide couplers with ultra-small electro-mechanical actuator," *Opt. Express*, vol. 18, no. 7, pp. 7031–7037, Mar. 2010.
[49] S. M. C. Abdulla *et al.*, "Tuning a racetrack ring resonator by an integrated dielectric MEMS cantilever," *Opt Express*, vol. 19, no. 17, pp. 15864–15878, 2011.
[50] L. J. Kauppinen *et al.*, "Micromechanically tuned ring resonator in silicon on insulator," *Opt. Lett.*, vol. 36, no. 7, p. 1047, Apr. 2011.
[51] C. Errando-Herranz, F. Niklaus, G. Stemme, and K. B. Gylfason, "A Low-power MEMS Tunable Photonic Ring Resonator for Reconfigurable Optical Networks," in *The 28th IEEE International Conference on Micro Electro Mechanical Systems (MEMS). Jan 2015*, 2015.
[52] C. Errando-Herranz, F. Niklaus, G. Stemme, and K. B. Gylfason, "A MEMS tunable photonic ring resonator with small footprint and large free spectral range," in *Solid-State Sensors, Actuators and Microsystems (TRANSDUCERS), 2015 Transducers-2015 18th International Conference on*, 2015, pp. 1001–1004.
[53] C. Errando-Herranz, F. Niklaus, G. Stemme, and K. B. Gylfason, "Low-power microelectromechanically tunable silicon photonic ring resonator add–drop filter," *Opt. Lett.*, vol. 40, no. 15, pp. 3556–3559, 2015.
[54] T. Ikeda and K. Hane, "A microelectromechanically tunable microring resonator composed of freestanding silicon photonic waveguide couplers," *Appl. Phys. Lett.*, vol. 102, no. 22, p. 221113+, Jun. 2013.
[55] T. Ikeda and K. Hane, "A tunable notch filter using microelectromechanical microring with gap-variable busline coupler," *Opt Express*, vol. 21, no. 19, pp. 22034–22042, 2013.
[56] H. M. Chu and K. Hane, "A Wide-Tuning Silicon Ring-Resonator Composed of Coupled Freestanding Waveguides," *Photonics Technol. Lett. IEEE*, vol. 26, no. 14, pp. 1411–1413, Jul. 2014.
[57] W. Jin, R. G. Polcawich, P. A. Morton, and J. E. Bowers, "Piezoelectrically tuned silicon nitride ring resonator," *Opt. Express*, vol. 26, no. 3, pp. 3174–3187, Feb. 2018.
[58] B. Dong, H. Tian, M. Zervas, T. J. Kippenberg, and S. A. Bhave, "PORT: A piezoelectric optical resonance tuner," in *2018 IEEE Micro Electro Mechanical Systems (MEMS)*, 2018, pp. 739–742.
[59] C. Castellan *et al.*, "Tuning the strain-induced resonance shift in silicon racetrack resonators by their orientation," *Opt. Express*, vol. 26, no. 4, pp. 4204–4218, Feb. 2018.
[60] H. Tian, B. Dong, M. Zervas, T. J. Kippenberg, and S. A. Bhave, "An unreleased MEMS actuated silicon nitride resonator with bidirectional tuning," in *2018 Conference on Lasers and Electro-Optics (CLEO)*, 2018, pp. 1–2.
[61] A. W. Elshaari *et al.*, "Strain-Tunable Quantum Integrated Photonics," *Nano Lett.*, vol. 18, no. 12, pp. 7969–7976, Dec. 2018.
[62] G. N. Nielson *et al.*, "Integrated wavelength-selective optical MEMS switching using ring resonator filters," *Photonics Technol. Lett. IEEE*, vol. 17, no. 6, pp. 1190–1192, Jun. 2005.
[63] C. Haffner *et al.*, "Sub-V Opto-Electro-Mechanical Switch," in *Conference on Lasers and Electro-Optics (2019), paper STh3H.4*, 2019, p. STh3H.4.
[64] Tianran Liu, Francesco Pagliano, René van Veldhoven, Vadim Pogoretskii, Yuqing Jiao, and Andrea Fiore, "Low-Voltage InP MEMS Optical Switch on Silicon," presented at the ECIO 2019, Gent, Belgium, 2019.
[65] M. W. Pruessner *et al.*, "InP-based optical waveguide MEMS switches with evanescent coupling mechanism," *J. Microelectromechanical Syst.*, vol. 14, no. 5, pp. 1070–1081, Oct. 2005.
[66] Z. K. Bishop *et al.*, "Electro-mechanical control of an on-chip optical beam splitter containing an embedded quantum emitter," *Opt. Lett.*, vol. 43, no. 9, pp. 2142–2145, May 2018.
[67] C. Papon *et al.*, "Nanomechanical single-photon routing," *Optica*, vol. 6, no. 4, pp. 524–530, Apr. 2019.
[68] Y. Akihama, Y. Kanamori, and K. Hane, "Ultra-small silicon waveguide coupler switch using gap-variable mechanism," *Opt. Express*, vol. 19, no. 24, pp. 23658–23663, Nov. 2011.
[69] K. Takahashi, Y. Kanamori, Y. Kokubun, and K. Hane, "A wavelength-selective add-drop switch using silicon microring resonator with a submicron-comb electrostatic actuator," *Opt Express*, vol. 16, no. 19, pp. 14421–14428, Sep. 2008.
[70] Y. Akihama and K. Hane, "Single and multiple optical switches that use freestanding silicon nanowire waveguide couplers," *Light Sci. Appl.*, vol. 1, no. 6, p. e16, Jun. 2012.
[71] T. Nagai and K. Hane, "Silicon photonic microelectromechanical switch using lateral adiabatic waveguide couplers," *Opt. Express*, vol. 26, no. 26, pp. 33906–33917, Dec. 2018.
[72] T. J. Seok, N. Quack, S. Han, and M. C. Wu, "50×50 Digital Silicon Photonic Switches with MEMS-Actuated Adiabatic Couplers," in *Optical Fiber Communication Conference (2015), paper M2B.4*, 2015, p. M2B.4.
[73] S. Han, T. J. Seok, N. Quack, B.-W. Yoo, and M. C. Wu, "Large-scale silicon photonic switches with movable directional couplers," *Optica*, vol. 2, no. 4, pp. 370–375, Apr. 2015.

[74] M. C. M. Lee and M. C. Wu, "MEMS-actuated microdisk resonators with variable power coupling ratios," *IEEE Photonics Technol. Lett.*, vol. 17, no. 5, pp. 1034–1036, May 2005.

[75] M.-C. M. Lee and M. C. Wu, "Tunable coupling regimes of silicon microdisk resonators using MEMS actuators," *Opt. Express*, vol. 14, no. 11, p. 4703, 2006.

[76] J. Yao, D. Leuenberger, M.-C. M. Lee, and M. C. Wu, "Silicon Microtoroidal Resonators With Integrated MEMS Tunable Coupler," *IEEE J. Sel. Top. Quantum Electron.*, vol. 13, no. 2, pp. 202–208, 2007.

[77] Z. Y. Li *et al.*, "A reconfigurable coupled optical resonators in photonic circuits for photon shutting," in *2017 19th International Conference on Solid-State Sensors, Actuators and Microsystems (TRANSDUCERS)*, 2017, pp. 619–620.

[78] X. Chew, G. Zhou, F. S. Chau, J. Deng, X. Tang, and Y. C. Loke, "Dynamic tuning of an optical resonator through MEMS-driven coupled photonic crystal nanocavities," *Opt. Lett.*, vol. 35, no. 15, pp. 2517–2519, Aug. 2010.

[79] S. Han, T. J. Seok, C.-K. Kim, R. S. Muller, and M. C. Wu, "Multicast silicon photonic MEMS switches with gap-adjustable directional couplers," *Opt. Express*, vol. 27, no. 13, p. 17561, Jun. 2019.

[80] T. Graziosi, H. Sattari, T. J. Seok, S. Han, M. C. Wu, and N. Quack, "Silicon photonic MEMS variable optical attenuator," in *MOEMS and Miniaturized Systems XVII*, 2018, vol. 10545, p. 105450H.

[81] T. Bakke, C. P. Tigges, and C. T. Sullivan, "1 /spl times/ 2 MOEMS switch based on silicon-on-insulator and polymeric waveguides," *Electron. Lett.*, vol. 38, no. 4, pp. 177–178, Feb. 2002.

[82] E. Bulgan, Y. Kanamori, and K. Hane, "Submicron silicon waveguide optical switch driven by microelectromechanical actuator," *Appl. Phys. Lett.*, vol. 92, no. 10, p. 101110, Mar. 2008.

[83] Timo Aalto, Mikko Harjanne, and Matteo Cherchi, "VTT's micron-scale silicon rib+strip waveguide platform," presented at the Proc.SPIE, 2016, vol. 9891.

[84] E. Aharon and D. Marom, "Fast switching via frustrated total internal reflection in silicon photonics MEMS-actuated waveguides," in *2019 International Conference on Optical MEMS and Nanophotonics*, Daejeon, Korea, 2019.

[85] R. Chatterjee and C. W. Wong, "Nanomechanical Proximity Perturbation for Switching in Silicon-Based Directional Couplers for High-Density Photonic Integrated Circuits," *J. Microelectromechanical Syst.*, vol. 19, no. 3, pp. 657–662, Jun. 2010.

[86] X. Sun *et al.*, "MEMS Tunable Hybrid Plasmonic-Si Waveguide," in *Optical Fiber Communication Conference (2017), paper Th2A.6*, 2017, p. Th2A.6.

[87] S. Han, T. J. Seok, N. Quack, B. Yoo, and M. C. Wu, "Monolithic 50×50 MEMS Silicon Photonic Switches with Microsecond Response Time," in *Optical Fiber Communication Conference*, San Francisco, California, 2014, p. M2K.2.

[88] T. J. Seok, N. Quack, S. Han, R. S. Muller, and M. C. Wu, "Large-scale broadband digital silicon photonic switches with vertical adiabatic couplers," *Optica*, vol. 3, no. 1, pp. 64–70, Jan. 2016.

[89] T. J. Seok, K. Kwon, J. Henriksson, J. Luo, and M. C. Wu, "Wafer-scale silicon photonic switches beyond die size limit," *Optica*, vol. 6, no. 4, p. 490, Apr. 2019.

[90] T. J. Seok, N. Quack, S. Han, W. Zhang, R. S. Muller, and M. C. Wu, "64×64 Low-loss and broadband digital silicon photonic MEMS switches," in *2015 European Conference on Optical Communication (ECOC)*, 2015, pp. 1–3.

[91] K. Kwon *et al.*, "128×128 Silicon Photonic MEMS Switch with Scalable Row/Column Addressing," in *Conference on Lasers and Electro-Optics*, San Jose, California, 2018, p. SF1A.4.

[92] N. Quack, T. J. Seok, S. Han, R. S. Muller, and M. C. Wu, "Scalable Row/Column Addressing of Silicon Photonic MEMS Switches," *IEEE Photonics Technol. Lett.*, vol. 28, no. 5, pp. 561–564, Mar. 2016.

[93] T. J. Seok *et al.*, "MEMS-Actuated 8×8 Silicon Photonic Wavelength-Selective Switches with 8 Wavelength Channels," in *Conference on Lasers and Electro-Optics (2018), paper STu4B.1*, 2018, p. STu4B.1.

[94] S. Han, T. J. Seok, K. Yu, N. Quack, R. S. Muller, and M. C. Wu, "Large-Scale Polarization-Insensitive Silicon Photonic MEMS Switches," *J. Light. Technol.*, vol. 36, no. 10, pp. 1824–1830, May 2018.

[95] S. Abe and K. Hane, "Variable-Gap Silicon Photonic Waveguide Coupler Switch With a Nanolatch Mechanism," *IEEE Photonics Technol. Lett.*, vol. 25, no. 7, pp. 675–677, Apr. 2013.

[96] M.-M. Lee, D. Hah, E. K. Lau, H. Toshiyoshi, and Ming Wu, "MEMS-actuated photonic crystal switches," *IEEE Photonics Technol. Lett.*, vol. 18, no. 2, pp. 358–360, Jan. 2006.

[97] S. M. C. Abdulla *et al.*, "Mechano-optical switching in a MEMS integrated photonic crystal slab waveguide," *2011 IEEE 24th Int. Conf. Micro Electro Mech. Syst.*, pp. 9–12, 2011.

[98] C. Errando-Herranz, M. Colangelo, S. Ahmed, J. Björk, and K. B. Gylfason, "MEMS tunable silicon photonic grating coupler for post-assembly optimization of fiber-to-chip coupling," in *2017 IEEE 30th International Conference on Micro Electro Mechanical Systems (MEMS)*, 2017, pp. 293–296.

[99] W. Yu *et al.*, "MEMS-Based Tunable Grating Coupler," *IEEE Photonics Technol. Lett.*, vol. 31, no. 2, pp. 161–164, Jan. 2019.

[100] C. Errando-Herranz, N. L. Thomas, and K. B. Gylfason, "Low-power optical beam steering by microelectromechanical waveguide gratings," *Opt. Lett.*, vol. 44, no. 4, pp. 855–858, Feb. 2019.

[101] M. Petruzzella *et al.*, "Electrically driven quantum light emission in electromechanically tuneable photonic crystal cavities," *Appl. Phys. Lett.*, vol. 111, no. 25, p. 251101, Dec. 2017.

[102] M. Petruzzella *et al.*, "Quantum photonic integrated circuits based on tunable dots and tunable cavities," *APL Photonics*, vol. 3, no. 10, p. 106103, Aug. 2018.

[103] M. Cotrufo, L. Midolo, Ž. Zobenica, M. Petruzzella, F. W. M. van Otten, and A. Fiore, "Nanomechanical control of optical field and quality factor in photonic crystal structures," *Phys. Rev. B*, vol. 97, no. 11, p. 115304, Mar. 2018.

[104] K. K. Tsia, S. Fathpour, and B. Jalali, "Electrical control of parametric processes in silicon waveguides," *Opt. Express*, vol. 16, no. 13, pp. 9838–9843, Jun. 2008.

[105] C. V. Poulton *et al.*, "Long-Range LiDAR and Free-Space Data Communication With High-Performance Optical Phased Arrays," *IEEE J. Sel. Top. Quantum Electron.*, vol. 25, no. 5, pp. 1–8, Sep. 2019.

[106] I. Zand, B. Abasahl, U. Khan, and W. Bogaerts, "Controlling parasitics in linear optical processors," in *IEEE Photonics Benelux Chapter/ Annual Symposium 2018*, Brussels, Belgium, 2018, pp. 1–4.

[107] J. Kim *et al.*, "Tunable Grating Couplers for Broadband Operation Using Thermo-Optic Effect in Silicon," *IEEE Photonics Technol. Lett.*, vol. 27, no. 21, pp. 2304–2307, Nov. 2015.

[108] S. Gao *et al.*, "Power-Efficient Thermal Optical Tunable Grating Coupler Based on Silicon Photonic Platform," *IEEE Photonics Technol. Lett.*, vol. 31, no. 7, pp. 537–540, Apr. 2019.

[109] C. P. Ho, Z. Zhao, Q. Li, S. Takagi, and M. Takenaka, "Tunable Grating Coupler by Thermal Actuation and Thermo-Optic Effect," *IEEE Photonics Technol. Lett.*, vol. 30, no. 17, pp. 1503–1506, Sep. 2018.

[110] N. Quack *et al.*, "Silicon Photonic MEMS: Exploiting Mechanics at the Nanoscale to Enhance Photonic Integrated Circuits," in *Optical Fiber Communication Conference (OFC) 2019*, San Diego, California, 2019, p. M2D.3.

[111] J. Jacobs *et al.*, "Die level release of silicon photonic MEMS," in *Optical MEMS and Nanophotonics (OMN), 2016 International Conference on*, 2016, pp. 1–2.

[112] L. Carroll *et al.*, "Photonic Packaging: Transforming Silicon Photonic Integrated Circuits into Photonic Devices," *Appl. Sci.*, vol. 6, no. 12, p. 426, Dec. 2016.

[113] P. Edinger, C. Errando-Herranz, and K. B. Gylfason, "Reducing Actuation Nonlinearity of MEMS Phase Shifters for Reconfigurable Photonic Circuits," in *Conference on Lasers and Electro-Optics (2019), paper SF2H.3*, 2019, p. SF2H.3.

[114] F. Niklaus, G. Stemme, J.-Q. Lu, and R. J. Gutmann, "Adhesive wafer bonding," *J. Appl. Phys.*, vol. 99, no. 3, p. 031101, Feb. 2006.

[115] X. Wang *et al.*, "Wafer-Level Vacuum Sealing by Transfer Bonding of Silicon Caps for Small Footprint and Ultra-Thin MEMS Packages," *J. Microelectromechanical Syst.*, vol. 28, no. 3, pp. 460–471, Jun. 2019.

[116] W. Spengen, R. Modlinski, R. Puers, and A. Jourdain, "Failure Mechanisms in MEMS/NEMS Devices," in *Springer Handbook of Nanotechnology*, B. Bhushan, Ed. Berlin, Heidelberg: Springer Berlin Heidelberg, 2007, pp. 1663–1684.

[117] A. L. Hartzell, M. G. Da Silva, and H. R. Shea, *MEMs reliability*. New York: Springer, 2011.

[118] N. T. Otterstrom, R. O. Behunin, E. A. Kittlaus, Z. Wang, and P. T. Rakich, "A silicon Brillouin laser," *Science*, vol. 360, no. 6393, pp. 1113–1116, Jun. 2018.
[119] S. Maity *et al.*, "Spectral Alignment of Single-Photon Emitters in Diamond using Strain Gradient," *Phys. Rev. Appl.*, vol. 10, no. 2, p. 024050, Aug. 2018.
[120] S. Meesala *et al.*, "Strain engineering of the silicon-vacancy center in diamond," *Phys. Rev. B*, vol. 97, no. 20, p. 205444, May 2018.
[121] Y.-I. Sohn *et al.*, "Controlling the coherence of a diamond spin qubit through its strain environment," *Nat. Commun.*, vol. 9, no. 1, p. 2012, May 2018.
[122] h2020morphic, "Home," *MORPHIC*. [Online]. Available: https://h2020morphic.eu/. [Accessed: 14-Jun-2019].
[123] C. Errando-Herranz *et al.,* "Dataset for "Review of MEMS for photonic integrated circuits": table for reported key figures ", *IEEE Dataport, 2019*. [Online]. Available: http://dx.doi.org/10.21227/fp2b-8b33. [Accessed: 16-Sep.-2019].

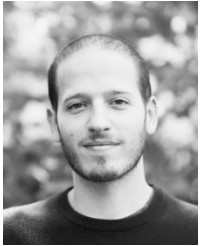
**Carlos Errando-Herranz** received the M.Sc. degree in automatic and electronic engineering from the Universitat Politècnica de València (UPV), Spain, in 2013, and the Ph.D. degree in Micro- and Nanosystems from the KTH Royal Institute of Technology, Stockholm, Sweden, in 2018. He is currently a postdoctoral researcher at the Quantum Nanophotonics group at KTH. His current research interests include integrated photonics, quantum photonics, and MEMS.

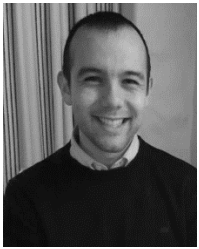
**Alain Yuji Takabayashi** received the M.Sc. degree in electrical engineering and information technology from the Eidgenössische Technische Hochschule Zürich (ETHZ), in 2017. He is currently pursuing the Ph.D. degree with the Institute of Microengineering at the Ecole Polytechnique Fédérale de Lausanne (EPFL), Switzerland. His current research is focused on the design, fabrication, and characterization of silicon photonic MEMS.

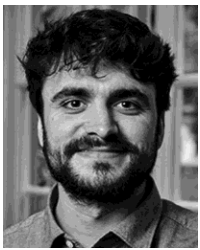
**Pierre Edinger** received the joint M.Sc. degree in micro- and nanotechnologies for integrated systems from INPG Phelma, France, the Politecnico di Torino, Italy, and EPFL, Switzerland, in 2017. He is currently pursuing the Ph.D. degree with the Department of Micro and Nanosystems, KTH Royal Institute of Technology, Stockholm, Sweden. His current research is focused on silicon photonics and MEMS.

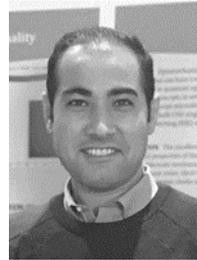
**Hamed Sattari** received the M.Sc. and Ph.D. degrees in Photonics-Telecommunication from University of Tabriz, Iran, in 2010 and 2014, respectively. After his PhD, he joined R&D unit of Nanotechnology Research Center (NANOTAM) at Bilkent University, Turkey, wherein he was designing plasmonic nano-antennas. In September 2016 he was awarded as a Swiss Government Excellence Scholarship holder for a Postdoctoral position at Q-Lab, Ecole Polytechnique Fédérale de Lausanne (EPFL). Since September 2017 he is involved in silicon photonics projects with goal of realizing MEMS-based reconfigurable components including switches, phase shifters, and variable optical attenuators.

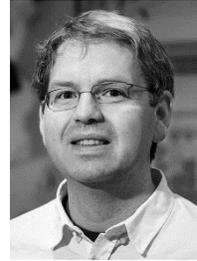
**Kristinn B. Gylfason** (M'99) received the B.Sc. and M.Sc. degrees in electrical engineering from the University of Iceland in 2001 and 2003, respectively, and the Ph.D. degree in electrical engineering from KTH in 2010. He also received the title of Docent in micro- and nanosystems from KTH in 2015. From 2003 to 2005, he was a Research Engineer with Lyfjathroun Biopharmaceuticals, Iceland. He is currently an Associate Professor, leading the team with a focus on photonic nanodevices for biomedical and communications applications. In 2005, he received the Steinmaur Foundation Nanotechnology Graduate Study Scholarship.

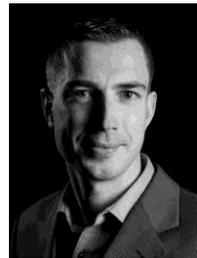
**Niels Quack** (S'05–M'11–SM'15) received the M. Sc. degree from Ecole Polytechnique Fédérale de Lausanne (EPFL), Switzerland, in 2005, and the Dr. Sc. degree from Eidgenössische Technische Hochschule Zürich (ETH), Switzerland, in 2010. From 2011 to 2015 he was Postdoctoral Researcher and Visiting Scholar at University of California, Berkeley, CA, USA, within the Integrated Photonics Laboratory at the Berkeley Sensor and Actuator Center. From 2014 to 2015 he was Senior MEMS Engineer with sercalo Microtechnology, Neuchâtel, Switzerland. He is currently SNSF Assistant Professor at Ecole Polytechnique Fédérale de Lausanne (EPFL), Lausanne, Switzerland. Research interests include Photonic Micro- and Nanosystems, with an emphasis on Diamond Photonics and Silicon Photonic MEMS. He is Senior Member of IEEE, Steering Committee Member of the IEEE International Conference on Optical MEMS and Nanophotonics (OMN), General Chair of the IEEE OMN 2018, Member of OSA and SPIE, and he has authored and co-authored more than 50 papers in leading technical journals and conferences.